\documentclass{ws-p9-75x6-50}

\begin{document}

\title{Behaviour of spin-half particles in curved space-time }

\author{Banibrata Mukhopadhyay}

\address{S.N.Bose National Centre for Basic Sciences, Block-JD,
Sector-III, Salt Lake, Calcutta-700098, India\\E-mail: bm@boson.bose.res.in
\footnote{Presently working in: 
Physical Research Laboratory, Navrangpura, Ahmedabad-380009, India
E-mail: bm@prl.ernet.in}}




\maketitle

\abstracts{
We study the behaviour of spin-half particles in curved space-time.
Since Dirac equation gives the dynamics of spin-half particles, we
mainly study the Dirac equation in Schwarzschild, Kerr,
Reissner-Nordstr\"om geometry. Due to the consideration of existence
of black hole in space-time (the curved space-time),
particles are influenced and equation will be modified.  
As a result the solution will be changed from that due to flat space.  
}

Here we study the interaction of particles having spin half
with a black hole. We are familiar
with Dirac equation in flat space by which we can investigate
the behaviour of spin-half particle. 
After writing the Dirac equation in curved space-time
particularly in Kerr or Kerr-Newman geometry
using Newman-Penrose formalism\cite{chan} general relativistic 
effect was introduced and the equation was generalised and modified.
As the space-time is asymptotically flat, far away from the 
black hole the modified Dirac equation in Kerr-Newman, 
Reissner-Nordstr\"om or Schwarzschild geometry and the interaction of particles
with space-time reduces into that of the flat space.

One of the most important solutions of Einstein's equation is that
of the spacetime around and inside an isolated black hole.
The spacetime at a large distance is flat and Minkowskian
where usual quantum mechanics is applicable, while the spacetime
closer to the singularity is so curved that no satisfactory
quantum field theory could be developed as yet. An intermediate
situation arises when a weak perturbation (due to, say, gravitational,
electromagnetic or Dirac waves) originating from
infinity impinges on a black hole, interacting with it.
The resulting wave is partially transmitted into the
black hole through the horizon and partially scatters off
to infinity. Particularly interesting is the fact that
whereas gravitational and electromagnetic radiations were found
to be amplified in some range of incoming frequencies,
Chandrasekhar\cite{chan} predicted that no such amplifications should
take place for Dirac waves because of the very nature of the
potential experienced by the incoming fields. However,
these later conclusions were drawn by Chandrasekhar using asymptotic
solutions. Here we revisit this important problem 
in a spatially complete manner to study the
nature of the Dirac spinors as a function of the
Kerr parameter, rest mass and frequency of incoming particle.

The separated Dirac equation in Kerr geometry (into radial
and angular part) is given by Chandrasekhar\cite{chan}.  
Following Mukhopadhyay \& Chakrabarti\cite{mc} we can solve
the equation in generalised Kerr-Newman geometry where the
basis vectors of the equation should be expressed in terms
of different coefficients of corresponding metric given by 

\begin{equation}
\begin{array}{rcl}
g^{ij}=\left(\begin{array}{cccc}\Sigma^2/\rho^2\Delta&0&0&2aMr/\rho^2\Delta\\
0&-\Delta/\rho^2&0&0\\0&0&-1/\rho^2&0\\2aMr/\rho^2\Delta&0&0&
-(\Delta-a^2sin^2\theta)/\rho^2{\Delta}sin^2\theta\end{array}\right)
\end{array}\label{eq:spa}
\end{equation}
where the different symbols have their usual meaning\cite{chan,bm1}. 
For simplicity we can choose the angular momentum of the black hole to
zero and space-time reduces to spherically symmetric Reissner-Nordstr\"om
geometry. Following Mukhopadhyay \& Chakrabarti\cite{mc} and Mukhopadhyay\cite{bm1}
we can reduce the radial Dirac equation into Schr\"odinger like equation by suitably
transforming the independent variable $r$ to $\hat{r}_*$ (which are
related to each other logarithmically\cite{mc,bm1,bm2}) and defining new spinor
fields as a certain linear combination of up and down spinors. 
Finally we solve the reduced radial equation
by {\it IWKB} approximation method\cite{mc,bm1} and get reflection and transmission
coefficients of Dirac particles as a function of rest mass and frequency of the incoming
Dirac particle. The angular equation was already solved earlier\cite{mc}.
Thus, combining the radial and angular solutions spatially complete solution of 
Dirac equation is obtained. 

If we consider the incoming Dirac particle as charged one (like electron, positron etc.)
then the covariance is appearing into the derivative of the equation not only due to the curvature of
the space (affine connections $\Gamma$) but also due to the electromagnetic interaction\cite{bm2}.
Thus the covariant derivative reduces as
\begin{equation}
\begin{array}{rcl}
D_\mu=\partial_\mu+iq_1A_\mu+q_2\Gamma_\mu
\end{array}\label{eq:spa}
\end{equation}
where, $q_1$ and $q_2$ are coupling constants. $A_\mu$ is the electromagnetic gauge field.

Another important issue in this perspective is
the super-radiance. In case of spherically symmetric Schwarzschild
geometry there is no question of super-radiance\cite{mc} but for the case of Kerr
or Kerr-Newman geometry there is a parameter region for which gravitational
potential diverges. But it is seen that potential varies as $\frac{1}{(r-|\alpha|)^3}$
($\alpha$ is a constant) close to the singular point.
As because potential has a strongly attractive branch 
practically no particle can come out from its field as a result super-radiation of Dirac
particle is impossible. Similar feature is seen in case of Reissner-Nordstr\"om geometry. 

The reflection and transmission coefficients of the Dirac particles
were found to distinguish strongly the solutions
of different rest masses and different energies.
The waves scattered off are distinctly different in different parameter regions.
In a way, therefore, black holes can act as a mass spectrograph!
Another interesting application of our method and solution would be to 
study the interactions of Hawking radiations in regions just outside the horizon.

\end{document}